\documentclass[a4paper,10pt,twoside,twocolumn]{article}
\usepackage{times}
\usepackage{bfcap}
\usepackage{graphicx}
\usepackage{fancyheadings}

\begin {document}

\pagestyle{fancyplain}
\lhead[\fancyplain{}{\thepage}]%
      {\fancyplain{}{\bfseries\leftmark \mdseries {{\sl Nature} {\bf 442}, 657-659 (10 August 2006)}}}
\rhead[\fancyplain{}{\bfseries\leftmark \mdseries {{\sl Nature} {\bf 442}, 657-659 (10 August 2006)}}]%
      {\fancyplain{}{\thepage}}
\cfoot{}

\noindent
{\LARGE\bf A probable stellar solution to the cosmological lithium discrepancy}\\

\noindent
{\large A.J.~Korn\footnote{Department of Astronomy and Space Physics, Uppsala University, Box 515, 75120 Uppsala, Sweden.}, F.~Grundahl\footnote{Department of Physics and Astronomy, University of \AA rhus, Ny Munkegade, 8000 \AA rhus C, Denmark.}, O.~Richard\footnote{3GRAAL-UMR5024/ISTEEM (CNRS), Universit\'{e}\ Montpellier II, Place E. Bataillon, 34095 Montpellier, France.}, P.S. Barklem$^1$, L.~Mashonkina\footnote{Institute of Astronomy, Russian Academy of Science, Pyatnitskaya 48, 119017 Moscow, Russia.}, R.~Collet$^1$, N. Piskunov$^1$ \& B.~Gustafsson$^1$}\\

\noindent
{\bf The measurement of the cosmic microwave background has
strongly constrained the cosmological parameters of the Universe\boldmath$^1$.
When the measured density of baryons (ordinary matter) is
combined with standard Big Bang nucleosynthesis calculations$^{2,3}$,
the amounts of hydrogen, helium and lithium produced shortly
after the Big Bang can be predicted with unprecedented
precision$^{1,4}$. The predicted primordial lithium abundance is a
factor of two to three higher than the value measured in the
atmospheres of old stars$^{5,6}$. With estimated errors of 10 to 25\,\%,
this cosmological lithium discrepancy seriously challenges our
understanding of stellar physics, Big Bang nucleosynthesis or
both. Certain modifications to nucleosynthesis have been proposed$^{7}$,
but found experimentally not to be viable$^{8}$. Diffusion
theory, however, predicts atmospheric abundances of stars to vary
with time$^{9}$, which offers a possible explanation of the discrepancy.
Here we report spectroscopic observations of stars in the metal-poor
globular cluster NGC\,6397 that reveal trends of atmospheric
abundance with evolutionary stage for various elements. These
element-specific trends are reproduced by stellar-evolution
models with diffusion and turbulent mixing$^{10}$. We thus conclude
that diffusion is predominantly responsible for the low apparent
stellar lithium abundance in the atmospheres of old stars by
transporting the lithium deep into the star.}

\noindent
Diffusive processes altering the elemental composition in stars
have been studied for decades$^{9,11}$. Evidence for their importance
comes from helioseismology$^{12}$ and the study of hot stars with peculiar
abundance patterns$^{13}$. Among solar-type stars, the effects of diffusion
are expected to be more pronounced in old, very metal-poor stars.
Given their greater age, diffusion has had more time to produce
sizeable effects than in younger stars like the Sun. Detailed
element-by-element predictions from models including effects of atomic
diffusion and radiative accelerations became available a few years
ago$^{14}$, but these early models produced strong abundance trends that
are incompatible with measurements of, in particular, the abundance
of lithium common among stars of the Galactic halo over a wide
range of metallicities (the so-called Spite plateau of lithium). However,
the recent inclusion of turbulent mixing$^{10}$ brings model predictions
into better agreement with observations.\\
According to the predictions from such model calculations, stars
leaving the main-sequence (turn-off stars) are expected to show the
largest variations relative to the composition of the gas from which
the stars originated. Giant stars, however, have deep surface convection
zones which erase most effects of diffusion and restore the
original composition. One notable exception is lithium which
disintegrates in layers with T $\geq$ 2.1 million K. The destruction of
lithium inside the star leads to a successive dilution of the surface
lithium because the convective envelope expands when the star
becomes a red giant. We performed spectroscopic observations
specifically to test these model predictions.\\
Globular-cluster stars have identical age and initial heavy-element
composition, and thus measured atmospheric abundance trends
with evolutionary stage are a signature of diffusion. We observed
18 stars along the evolutionary sequence of NGC\,6397 with the
multi-object spectrograph FLAMES-UVES on the 8.2-m Very Large
Telescope in Chile -- five stars close to the main-sequence turn-off
point (TOP), two stars in the middle of the subgiant branch (SGB),
five stars at the base of the red-giant branch (bRGB) and six red giants
(RGB) -- that represent specific stages in stellar evolution (Fig.~2). Even on 8-m-class telescopes, these observations
are still challenging, requiring total integration times of 2-12 hours
per star to obtain homogeneous data with high signal-to-noise ratios
at high spectral resolution.\\

\Large
\small
\begin{table*}[t]
\begin{center}
{\bf Mean stellar parameters, iron abundances and photometric
quantities of the four groups of stars\vspace*{2mm}}
\begin{tabular}{lccccc}
\hline
Group & No. of star & $T_{\rm eff}$ (K) & log [$g$ (cm\,s$^{-2}$)] & $\log [\varepsilon$\,(Fe)] & $\xi$ (km\,s$^{-1}$)\\
\hline
TOP & 5 & 6254 & 3.89 & 5.23 $\pm$ 0.04 & 2.00 \\
SGB & 2 & 5805 & 3.58 & 5.27 $\pm$ 0.05 & 1.75 \\
bRGB & 5 & 5456 & 3.37 & 5.33 $\pm$ 0.03 & 1.73 \\
RGB & 6 & 5130 & 2.56 & 5.39 $\pm$ 0.02 & 1.60 \\
Sun & 1 & 5777 & 4.44 & 7.51 & 1.00 \\
\hline
 & \multicolumn{2}{l}{$\Delta$\,$T_{\rm eff}$ (TOP\,$-$\,RGB)} (K)\ \ & $\Delta$\,$\log g$ (TOP\,$-$\,RGB) & \multicolumn{2}{l}{$\Delta$\,$\log \varepsilon$\,(Fe) (TOP\,$-$\,RGB)}\\
\hline
Spectroscopy & \multicolumn{2}{c}{1124} & 1.33 & \multicolumn{2}{c}{0.16 $\pm$ 0.05}\\
Str\"{o}mgren, ($v-y$) & \multicolumn{2}{c}{1108} & 1.38 & \multicolumn{2}{c}{--}\\
Broad-band, ($V-I$) & \multicolumn{2}{c}{1070} & 1.38 & \multicolumn{2}{c}{--}\\
\hline
\end{tabular}
\end{center}
\caption{Mean stellar parameters of the four groups of stars in the following evolutionary stages:
TOP, SGB, bRGB and RGB (see text). The FLAMES-UVES spectra cover the spectral range
from 4800\,-\,6800 \AA, have R = $\lambda/\Delta\lambda$ = 48\,000 (where $\lambda$ is wavelength) and signal-to-noise
ratios in excess of 80:1 per pixel. The analyses are fully spectroscopic and line-by-line
differential to the Sun. Typical errors on $T_{\rm eff}$, log\,$g$ and the microturbulence $\xi$ for individual
stars are, respectively, 150\,K, 0.15 and 0.2 km\,s$^{-1}$. The errors in log[$\varepsilon$(Fe)] = log($N_{\rm Fe}/
N_{\rm H}$) + 12 are the combined values of the line-to-line scatter of Fe\,{\sc i} and Fe\,{\sc ii} for the individual
stars propagated into the mean value for the group. Between 20 and 40 Fe\,{\sc i} and Fe\,{\sc ii} lines
were measured by means of profile fits. The stellar parameters of the TOP and SGB group
have not been corrected for helium diffusion, which would result in higher log\,$g$ values
(+0.05; see text). Below, the spectroscopic results are compared with the photometry
(Str\"{o}mgren and broad-band indices calibrated on the infrared-flux method$^{28}$, see also
Table 2) obtained with the Danish 1.54-m telescope on La Silla. In both cases,
$\Delta$ log\,$g$ is based on the magnitude difference in $V$.}
\end{table*}

\Large
\small
{\normalsize These observational challenges explain in part why there have been
surprisingly few studies looking for abundance trends between
unevolved and evolved stars in globular clusters. One study$^{15}$
found lower abundances for subgiants in the very metal-poor
globular cluster M\,92 than accepted literature values indicate for
giant stars. However, the low signal-to-noise data did not allow firm
conclusions. A different study$^{16}$ examined TOP stars and subgiant
stars in NGC\,6397. It concentrated on elements showing suspected
nuclear processing (oxygen, sodium, magnesium and aluminium).
The analysis was based on standard models of stellar evolution and
spectroscopic analysis (assuming equilibrium) and did not indicate
significant differences in iron abundances. This result critically
depends on the above-mentioned assumptions, which yield stellar
parameters (effective temperatures) unsupported by photometry.
The only previous study attempting a homogeneous analysis of red-giant
and dwarf stars$^{17}$ targeted M\,13, a globular cluster a factor of
three more metal-rich than NGC\,6397. The highest-gravity object (a
subgiant with log\,$g$\,=\,3.8) indeed shows an iron abundance 41\,\%
(0.23 in log abundance) below the average of the other 24 more evolved
cluster members analysed. This was not interpreted as an
indication of departures from a uniform cluster composition as
caused by diffusion.\\
To derive stellar parameters and elemental abundances from our
observations, we employed well-established spectroscopic diagnostics
(in one dimension) with a high level of modelling realism in the
line formation (non-equilibrium where necessary). The profile of
the Balmer line of hydrogen (H$\alpha$, at 656 nm) was used to estimate the
effective temperatures ($T_{\rm eff}$)$^{18,19}$, while the ionization equilibrium of
iron (treated in non-equilibrium) was used to determine the stellar
surface gravities (log\,$g$)$^{20}$. The mean stellar parameters for the four
groups of stars are given in Table~1.\\
\begin{figure*}[!t]\centering
\includegraphics[width=0.72\textwidth,clip=]{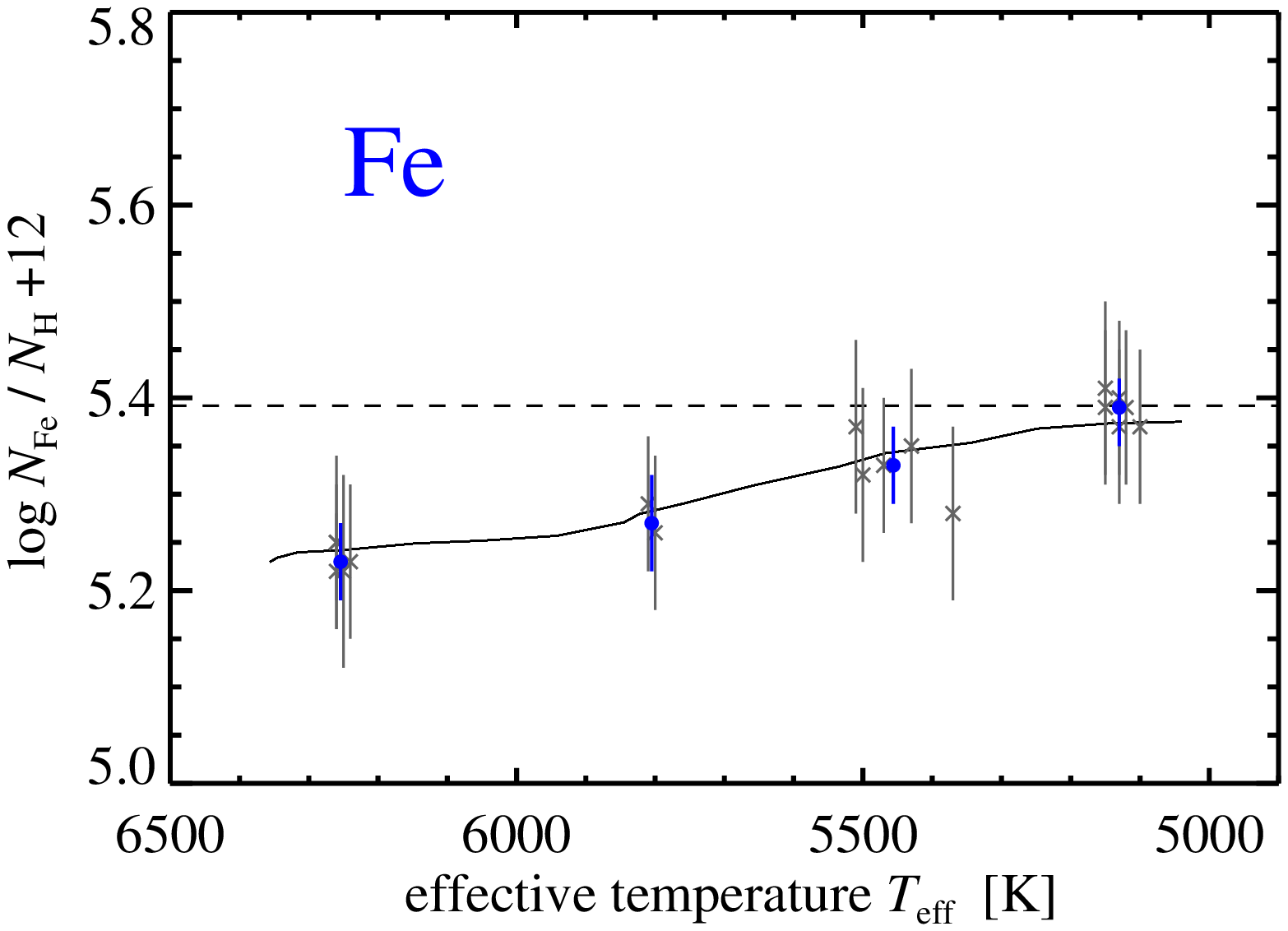}
\includegraphics[width=0.72\textwidth,clip=]{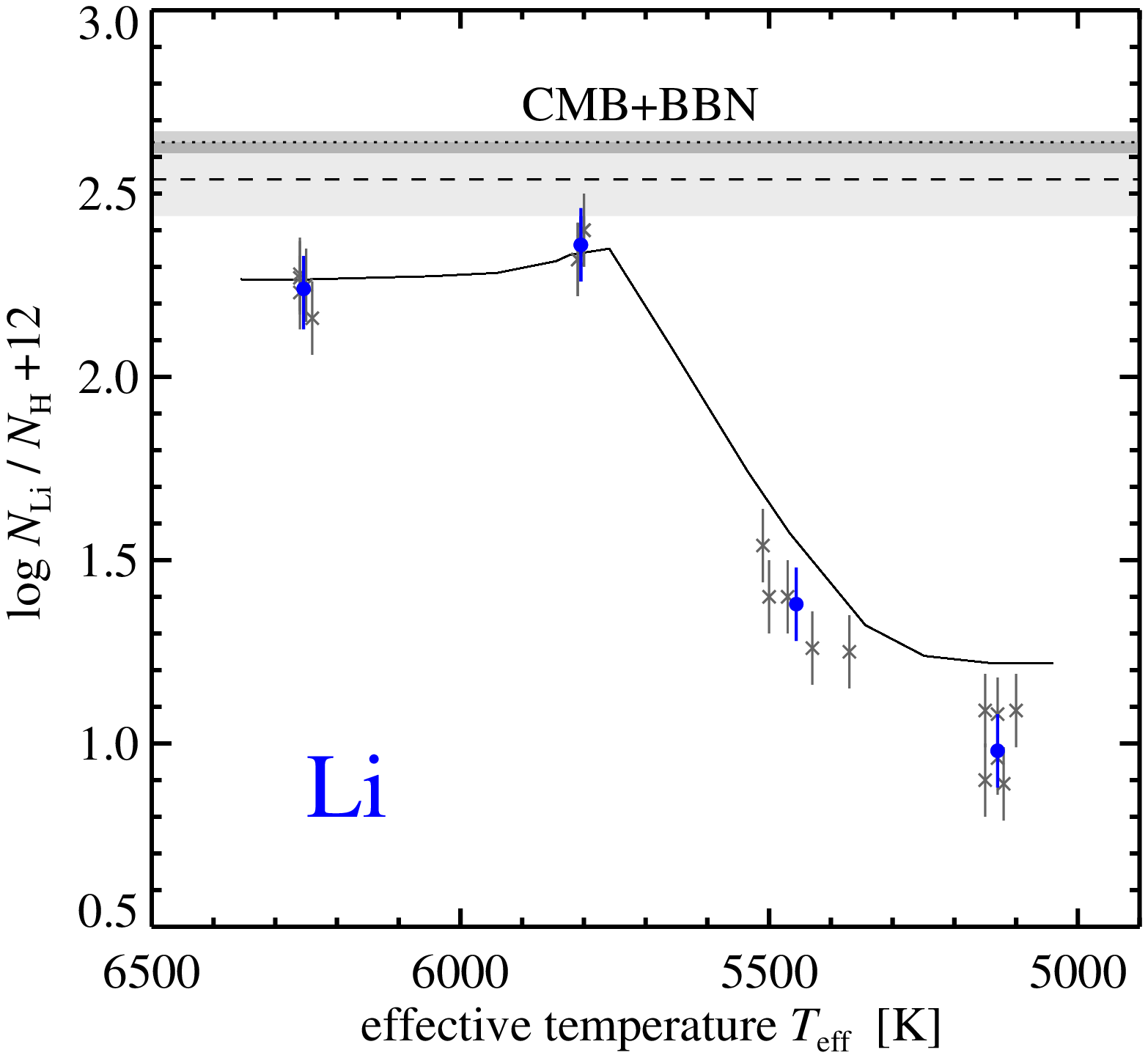}
\caption{{\bf Trends of iron and lithium as a function of the effective
temperatures of the observed stars compared to the model predictions.}
The grey crosses are the individual measurements, while the bullets are the
group averages. The solid lines are the predictions of the diffusion model,
with the original abundance given by the dashed line. In {\em b}, the grey-shaded
area around the dotted line indicates the 1$\sigma$ confidence interval of
CMB + BBN$^1$: log[$\varepsilon$(Li)] = log ($N_{\rm Li}/N_{\rm H}$) + 12 = 2.64 $\pm$ 0.03. In {\em a}, iron is
treated in non-equilibrium$^{20}$ (non-LTE), while in {\em b}, the equilibrium (LTE)
lithium abundances are plotted, because the combined effect of 3D and
non-LTE corrections was found to be very small$^{29}$. For iron, the error bars
are the line-to-line scatter of Fe\,{\sc i} and Fe\,{\sc ii} (propagated into the mean for the
group averages), whereas for the absolute lithium abundances 0.10 is
adopted. The 1$\sigma$ confidence interval around the inferred primordial lithium
abundance (log[$\varepsilon$(Li)] = 2.54 $\pm$ 0.10) is indicated by the light-grey area.We
attribute the modelling shortcomings with respect to lithium in the bRGB
and RGB stars to the known need for extra mixing$^{30}$, which is not considered
in the diffusion model.}
\end{figure*}
An independent check of the reliability of the stellar-parameter determination is obtained based on photometry, exploiting the
luminosity difference of the stars. These data are also given in Table~1. As can be seen, the spectroscopic $\Delta$\,$T_{\rm eff}$ (TOP\,$-$\,RGB) (that is, $T_{\rm eff}$ (TOP) $-$
$T_{\rm eff}$ (RGB)) is well-matched by both the Str\"{o}mgren index ($v-y$) and the
broad-band index ($V-I$), the photometry indicating a $\Delta T_{\rm eff}$ only
20-50\,K (2-5\,\%) lower. The spectroscopic $\Delta$\,$\log g$ (TOP\,$-$\,RGB) is only 0.05
below that determined from the photometry. This is excellent agreement
for two fully independent methods and we thus consider the
stellar-parameter differences to be well constrained. On the basis of
these differences, relative abundance trends can be scrutinized.\\
The best-determined abundance (in terms of number of spectral
lines used) is that of iron. We find the abundance to differ by 45\,\%
(0.16 in log abundance), that is, the TOP stars have the lowest
abundance which successively rises towards the RGB stars (Fig.~1a).
Propagating the line-to-line scatter for all iron lines of each
star into the mean values given in Table 1 and further into the
abundance difference between TOP and RGB stars, the difference
$\Delta$\,$\log \varepsilon$\,(Fe) (TOP\,$-$\,RGB) is significant at the 3.2$\sigma$ level. Calcium
and titanium also show trends (Fig.~3), but
these are less pronounced.\\
We have compared these abundance trends with various diffusion
model predictions and found that a model with one particular value
of the turbulent-mixing efficiency$^{10}$ is capable of fitting the observations
of these heavy elements well. This model (model T6.0, with a
parametric description of turbulent mixing using an isotropic
turbulent diffusion coefficient 400 times larger than the atomic
diffusion coefficient for helium at log\,$T$ = 6.0 varying with density
as $\rho^{-3}$ (ref. 10), and computed for the metallicity of the RGB stars)
clearly predicts a steeper trend for iron than for calcium or titanium,
in agreement with the observations. The simultaneous element-specific
reproduction of the observations thus lends strong support
for the diffusion interpretation of these trends and constrains the
level of turbulent mixing.\\
We emphasize that it is not possible to remove all trends simultaneously
by adjusting the stellar parameters: the neutral species Ca\,{\sc i}
and Fe\,{\sc i} (both affected by non-equilibrium) require a $T_{\rm eff}$ correction
of roughly 100\,K and 200\,K respectively, while the ionized species Ti\,{\sc ii}
and Fe\,{\sc ii} (formed in equilibrium) would require a change in log\,$g$
of 0.15 and 0.33, respectively. Considering the agreement between spectroscopy and photometry, these corrections to the stellar parameters
are large and incompatible with one another.
We also investigate the effect that the three-dimensional (3D)
hydrodynamic nature of stellar atmospheres$^{21}$ might have on the
observed trends. Using 3D TOP and RGB models available to us,
we find that the trends inferred from weak Fe\,{\sc ii} and Ti\,{\sc ii} lines in
one dimension require small 3D corrections only. For iron, the trend
would even be slightly steeper. As the stellar-parameter differences
are well determined, analyses based on 3D hydrodynamic models
would reach much the same conclusion about the abundance trends.\\
In the light of the identified diffusion signature, we should
consider the structural effect that helium diffusion has on the
atmosphere and, therefore, on the spectroscopic analysis. The T6.0
model predicts the He/H ratio in the TOP stars to be decreased by
nearly 50\% from the original value, 40\% in the SGB stars, whereas the
original ratio is almost restored in the bRGB and RGB stars. Helium
settling of this extent changes the mean molecular weight in the atmosphere appreciably, an effect which can be mapped as a shift in
gravity$^{22}$. The accompanying shift in log\,$g$ amounts to +0.05 for both
groups, bringing the spectroscopic $\Delta$\,$\log g$ (TOP\,$-$\,RGB) into perfect agreement
with the photometry-based value. This correction also
improves the agreement of the stellar parameters with the isochrone
(Fig.~2).\\
Lithium is ionized easily and therefore behaves much like helium
in terms of settling. As seen from Fig.~1b, both the TOP stars and
the SGB stars sample the lithium plateau. The observations support
the T6.0 model predictions, with a slight upturn towards cooler
temperatures before the convection zone encompasses lithium-free
layers and dilution sets in. A higher value of lithium among halo
subgiants than among halo dwarf stars was recently reported$^{23}$ and
qualitatively interpreted as a signature of atomic diffusion counterbalanced
by gravity-wave-induced mixing. In quantitative terms,
the diffusion model we use predicts the original lithium abundance
to be 78\% (0.25 in log abundance) higher than the mean value
measured in the TOP and SGB stars. Therefore, we infer a
primordial lithium abundance which agrees with the cosmic microwave
background plus Big Bang nucleosynthesis (CMB + BBN)
value$^1$ within the mutual 1$\sigma$ error bars. Given the uncertainties
associated with both values, the cosmological lithium discrepancy is
thus largely removed. This finding restores confidence in standard
BBN, quantitative spectroscopy and sophisticated stellar evolution
models.\\
In extrapolating to the primordial lithium abundance, we do not
consider corrections related to the Galactic chemical evolution of
lithium. Such corrections are model-dependent and rather uncertain.
It is plausible that some fraction of the halo gas was processed
through Population III stars$^{24}$ (lowering the lithium abundance) and
that cosmic rays interacting with the interstellar medium have
produced some lithium$^{6}$. At present it is, however, unclear to
which extent (and even in which direction) the stellar lithium
abundances should be adjusted.\\
While helium diffusion has been invoked to lower globular-cluster
ages to values which meet the cosmological age constraint, metal
diffusion has rarely been considered. The models including atomic
diffusion, radiative accelerations and turbulent mixing for all
elements will result in moderately different absolute and relative
ages. Using similar models$^{25}$, the analysis of M\,92 has led to an age
estimate compatible with Wilkinson Microwave Anisotropy Probe
(WMAP$^1$) results. It is clear, however, that stellar-evolution models
need to be developed further towards physical self-consistency with
respect to all relevant processes, including hydrodynamics and
rotation. In particular, the physical mechanism causing turbulent
mixing should be identified. Promising work$^{26}$ points towards the
importance of internal gravity waves in transporting angular
momentum and mixing stellar interiors.\\
Regarding unevolved metal-poor stars as tracers of Galactic cosmochemistry,
heavy-element abundance ratios (for example, Mg/Fe) are
less affected by diffusion than ratios with respect to hydrogen (for
example, Fe/H). However, the effects are non-negligible and must be
taken into account to reach the highest accuracy. Care should be
taken especially when comparing abundances from giant and dwarf
stars. In this respect, one exceptional pair is the two most metal-poor
stars known$^{27}$, whose abundance signatures will have to be
reinvestigated in the light of diffusion.\\

\noindent
{\bf Received 28 March; accepted 14 June 2006.}\\

\noindent
1. Spergel, D. N. et al. Wilkinson Microwave Anisotropy Probe (WMAP) three
year results: implications for cosmology. Astrophys. J. (submitted); preprint at
http:// arxiv.org/abs/astro-ph/0603449l (2006).\\
2. Wagoner, R. V., Fowler, W. A. \& Hoyle, F. On the synthesis of elements at very
high temperatures. Astrophys. J. 148, 3–-49 (1967).\\
3. Burles, S., Nollett, K. M. \& Turner, M. S. Big bang nucleosynthesis predictions
for precision cosmology. Astrophys. J. 552, L1–-L5 (2001).\\
4. Cyburt, R. H., Fields, B. D. \& Olive, K. A. Primordial nucleosynthesis in light of
WMAP. Phys. Lett. B 567, 227–-234 (2003).\\
5. Spite, M. \& Spite, F. Lithium abundance at the formation of the galaxy. Nature
297, 483–-485 (1982).\\
6. Ryan, S. G., Norris, J. E. \& Beers, T. C. The Spite lithium plateau: ultrathin but
postprimordial. Astrophys. J. 523, 654–-677 (1999).\\
7. Coc, A., Vangioni-Flam, E., Descouvemont, P., Adahchour, A. \& Angulo, C.
Updated big bang nucleosynthesis compared with Wilkinson Microwave
Anisotropy Probe observations and the abundance of light elements. Astrophys.
J. 600, 544–-552 (2004).\\
8. Angulo, C. et al. The $^7$Be(d,p)2$\alpha$ cross section at big bang energies and the
primordial $^7$Li abundance. Astrophys. J. 630, L105–-L108 (2005).\\
9. Aller, L. H. \& Chapman, S. Diffusion in the sun. Astrophys. J. 132, 461–-472 (1960).\\
10. Richard, O., Michaud, G. \& Richer, J. Implications of WMAP observations
on Li abundance and stellar evolution models. Astrophys. J. 619, 538–-548
(2005).\\
11. Michaud, G., Fontaine, G. \& Beaudet, G. The lithium abundance: constraints on
stellar evolution. Astrophys. J. 282, 206–-213 (1984).\\
12. Guzik, J. A. \& Cox, A. N. On the sensitivity of high-degree p-mode frequencies to
the solar convection zone helium abundance. Astrophys. J. 386, 729–-733 (1992).\\
13. Richer, J., Michaud, G. \& Turcotte, S. The evolution of AMFM stars, abundance
anomalies, and turbulent transport. Astrophys. J. 529, 338–-356 (2000).\\
14. Richard, O., Michaud, G. \& Richer, J. Models of metal-poor stars with
gravitational settling and radiative accelerations. III. Metallicity dependence.
Astrophys. J. 580, 1100–-1117 (2002).\\
15. King, J. R., Stephens, A., Boesgaard, A. M. \& Deliyannis, C. Keck HIRES
spectroscopy of M92 subgiants -- surprising abundances near the turnoff.
Astron. J. 115, 666–-684 (1998).\\
16. Gratton, R. G. et al. The O-Na and Mg-Al anticorrelations in turn-off and early
subgiants in globular clusters. Astron. Astrophys. 369, 87–-98 (2001).\\
17. Cohen, J. G. \& Mele´ndez, J. Abundances in a large sample of stars in M3 and
M13. Astron. J. 129, 303–-329 (2005).\\
18. Fuhrmann, K., Axer, M. \& Gehren, T. Balmer lines in cool dwarf stars.
I. Basic influence of atmospheric models. Astron. Astrophys. 271, 451–-462 (1993).\\
19. Barklem, P. S., Piskunov, N. \& O'Mara, B. J. Self-broadening in Balmer line
wing formation in stellar atmospheres. Astron. Astrophys. 363, 1091–-1105
(2000).\\
20. Korn, A. J., Shi, J. \& Gehren, T. Kinetic equilibrium of iron in the atmospheres of
cool stars. III. The ionization equilibrium of selected reference stars. Astron.
Astrophys. 407, 691–-703 (2003).\\
21. Stein, R. F. \& Nordlund, \AA. Simulations of solar granulation. I. General
properties. Astrophys. J. 499, 914–-933 (1998).\\
22. Str\"{o}mgren, B., Gustafsson, B. \& Olsen, E. H. Evidence of helium abundance
differences between the Hyades stars and field stars, and between Hyades
stars and Coma cluster stars. Astron. Soc. Pacif. 94, 5–-15 (1982).\\
23. Charbonnel, C. \& Primas, F. The lithium content of the galactic halo stars.
Astron. Astrophys. 442, 961–-992 (2005).\\
24. Piau, L. et al. From first stars to the Spite plateau: a possible reconciliation of
halo stars observations with predictions from big bang nucleosynthesis.
Astrophys. J. (submitted); preprint at http://arxiv.org/abs/astro-ph/0603553l (2006).\\
25. VandenBerg, D. A., Richard, O., Michaud, G. \& Richer, J. Models of metal-poor
stars with gravitational settling and radiative accelerations. II. The age of the
oldest stars. Astrophys. J. 571, 487–-500 (2002).\\
26. Charbonnel, C. \& Talon, S. Influence of gravity waves on the internal rotation
and Li abundance of solar-type stars. Science 309, 2189–-2191 (2005).\\
27. Frebel, A. et al. Nucleosynthetic signatures of the first stars. Nature 434,
871–-873 (2005).\\
28. Alonso, A., Arribas, S. \& Martinez-Roger, C. The effective temperature scale of
giant stars (F0–-K5). II. Empirical calibration of $T_{\rm eff}$ versus colours and [Fe/H].
Astron. Astrophys. Suppl. 140, 261–-277 (1999).\\
29. Barklem, P. S., Belyaev, A. K. \& Asplund, M. Inelastic H + Li and H$^-$ + Li$^+$
collisions and non-LTE Li\,{\sc i} line formation in stellar atmospheres. Astron.
Astrophys. 409, L1–-L4 (2003).\\
30. Charbonnel, C. A consistent explanation for $^{12}$C/$^{13}$C, $^{7}$Li and $^{3}$He anomalies in
red giant stars. Astrophys. J. 453, L41–-L44 (1995).\\

\noindent
{\bf Supplementary Information} is linked to the online version of the paper at
www.nature.com/nature.\\

\noindent
{\bf Acknowledgements} A.J.K. acknowledges a research fellowship by the
Leopoldina Foundation, Germany. O.R. thanks the Centre Informatique National
de l'Enseignement Sup\'{e}rieur (CINES) and the R\'{e}seau Qu\'{e}b\'{e}cois de Calcul de
Haute Performance (RQCHP) for providing the computational resources
required for this work. F.G. acknowledges financial support from the Instrument
Center for Danish Astrophysics (IDA). L.M. acknowledges support through the
Presidium RAS Programme `Origin and evolution of stars and the Galaxy'. The
Uppsala group of authors acknowledges support from the Swedish Research
Council. We thank A. Alonso and I. Ramirez for providing colour–temperature
relations specific to this project.\\

\noindent
{\bf Author Information} Reprints and permissions information is available at
npg.nature.com/reprintsand\-permissions. The authors declare no competing
financial interests. Correspondence and requests for materials should be
addressed to A.J.K. (akorn@astro.uu.se).}

\newpage
\noindent
{\LARGE\bf Supplementary information}

\begin{figure*}[!t]\centering
\includegraphics[width=0.84\textwidth,clip=]{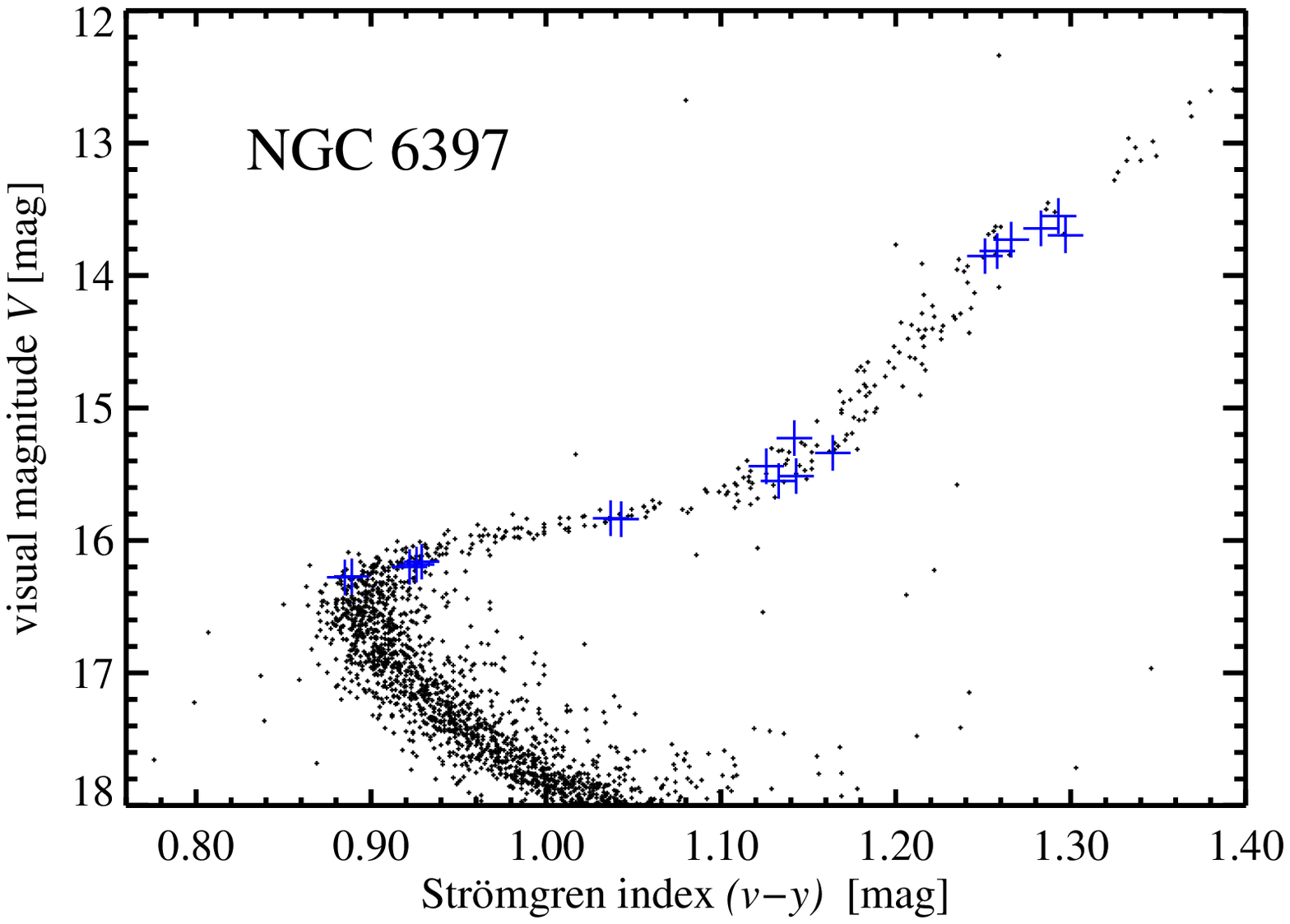}
\hspace*{-5mm}
\includegraphics[width=0.82\textwidth,clip=]{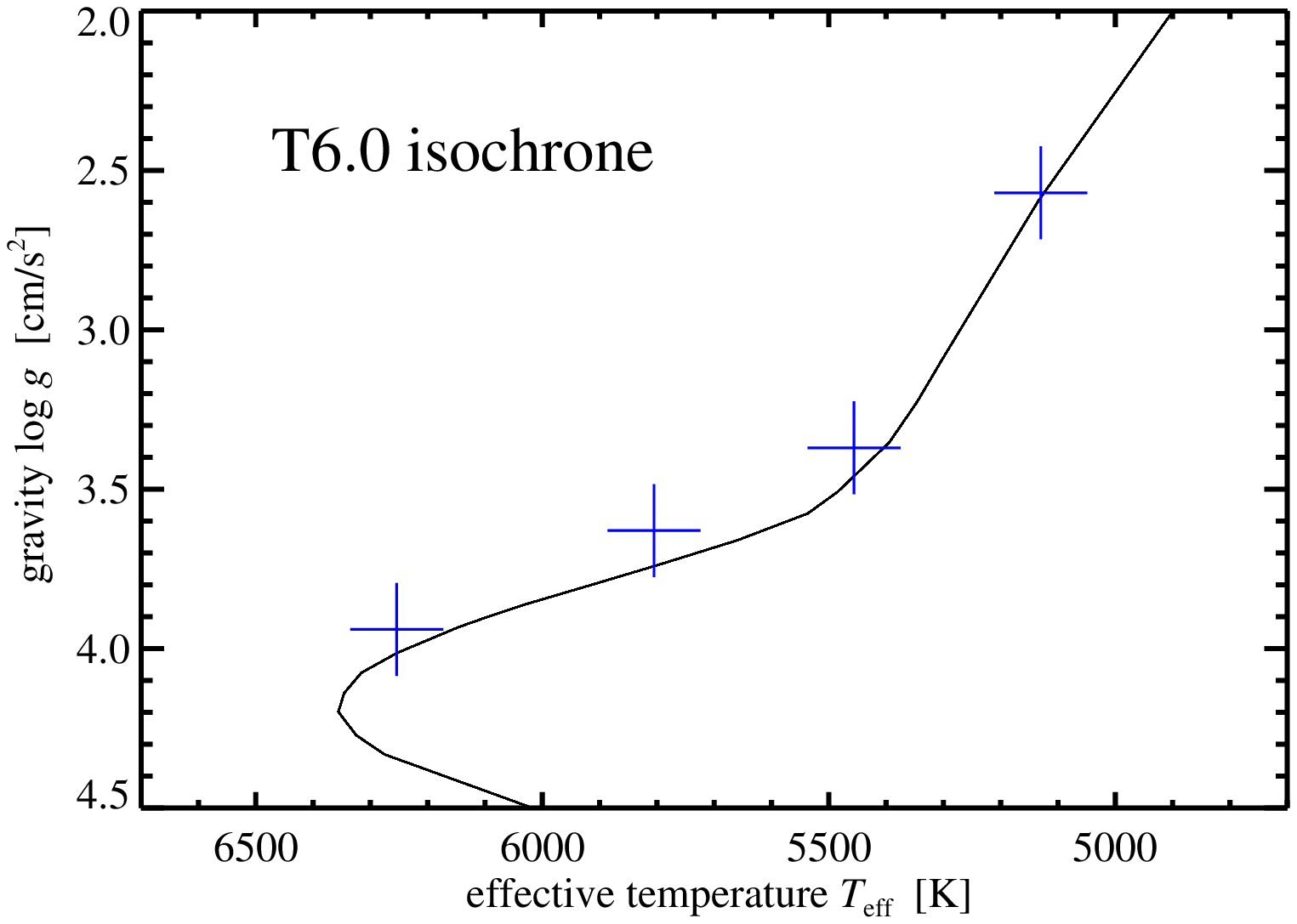}
\caption{{\bf Loci of the observed stars in the observational and physical parameter space.}
{\em Top:} Colour-magnitude diagram of NGC\,6397 clearly showing the main sequence, the subgiant branch and the red-giant branch. The observed stars are indicated by the blue crosses.
{\em Bottom:} Comparison of the spectroscopic stellar parameters (corrected for helium diffusion, see text) with a 13.5\,Gyr isochrone constructed from the diffusion model T6.0. The agreement is satisfactory indicating that the absolute stellar parameters meet the cosmological age constraint.}
\end{figure*}

\begin{figure*}[!t]\centering
\includegraphics[width=0.80\textwidth,clip=]{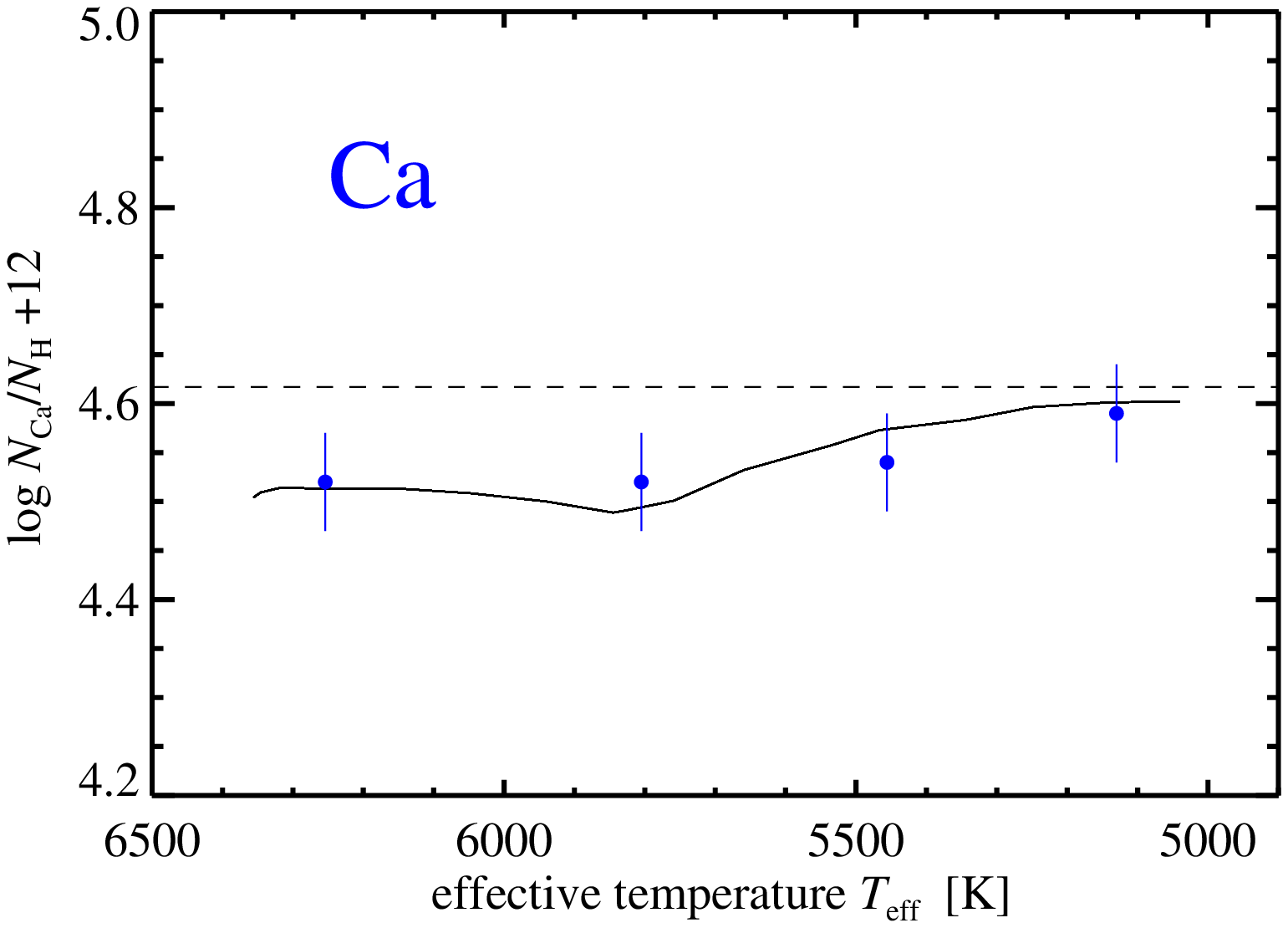}
\includegraphics[width=0.80\textwidth,clip=]{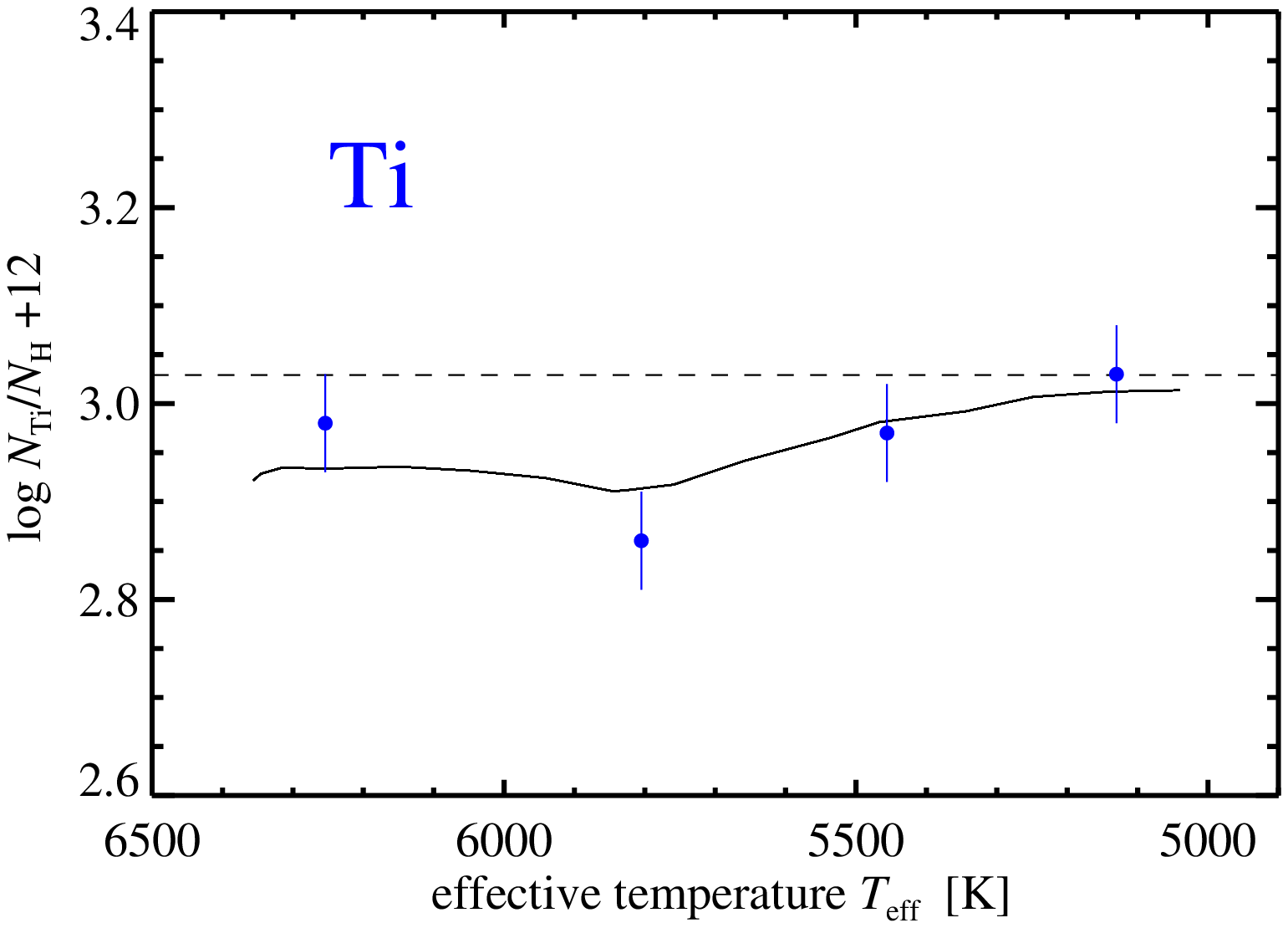}
\caption{{\bf Trends of calcium and titanium as a function of the effective temperatures of the observed stars compared to the model predictions.} As weak lines were used, the analyses were done on the mean spectra for each group of stars using the mean stellar parameters given in Table 1. The calcium abundances ({\em top}) were determined in non-LTE from three neutral lines (at 612.2, 616.2 and 643.9\,nm) which have line strengths between 30\,m\AA\ (Ca\,{\sc i} 612.2 in the TOP stars) and 92\,m\AA\ (Ca\,{\sc i} 616.2 in the RGB stars). The titanium abundances ({\em bottom}) were determined from the only available ionized line (Ti\,{\sc ii} 522.6\,nm) which has line strengths between 22\,m\AA\ (TOP) and 67\,m\AA\ (RGB). For the relative abundance trends errors of 0.05 are adopted. The observed trends (markedly shallower than for iron) are in good agreement with the diffusion predictions (full-drawn line; the dashed line indicates the original abundances used in the model).
Trends were observed for other elements (e.g.~scandium, magnesium), but no model predictions exist for some (e.g.~scandium) and others (e.g.~magnesium, aluminium) suffer from suspected nuclear processing in globular clusters$^{16}$ and are thus not suitable for isolating the diffusion signature.}
\end{figure*}

\Large
\small
\begin{table*}[t]
\begin{center}
{\bf Spectroscopic versus photometric effective temperatures of the four groups of stars observed\vspace*{2mm}}
\begin{tabular}{lclll}
\hline
Group & No. of stars & \,$T_{\rm eff}$(spec) (K) & \,$T_{\rm eff}$($v-y$) (K) & \,$T_{\rm eff}$($V-I$) (K)\\
\hline
TOP & 5 & 6254 $\pm$ \,\ 9 & 6240 $\pm$ 21 & 6133 $\pm$ 31 \\
SGB & 2 & 5805 $\pm$ \,\ 7 & 5824 $\pm$ \,\ 6 & 5688 $\pm$ \,\ 8 \\
bRGB & 5 & 5456 $\pm$ 57 & 5408 $\pm$ 37 & 5318 $\pm$ 29 \\
RGB & 6 & 5130 $\pm$ 19 & 5132 $\pm$ 22 & 5063 $\pm$ 20 \\
\hline
\end{tabular}
\end{center}
\caption{The photometric indices are calibrated on the infrared-flux method$^{28}$. The agreement between spectroscopy and ($v-y$) is excellent, whereas ($V-I$) is offset towards cooler effective temperatures by roughly 100\,K. However, for the identification of the diffusion signature the effective-temperature difference is the most relevant quantity. The effective-temperature difference between TOP and RGB stars in ($V-I$) is lower than in the other two cases (by roughly 50\,K). Because of this, a slightly larger abundance difference would result when referring to the ($V-I$) temperature scale. The conclusion is that three prominent effective-temperature scales (H$\alpha$ spectroscopy, Str\"{o}mgren and broad-band photometry) reveal significant abundance differences between the TOP and RGB stars of NGC\,6397.}
\end{table*}

\Large
\small
\end{document}